\documentclass[%
preprint,
 amsmath,amssymb,
 aps,
]{revtex4-1}

\usepackage{graphicx} 
\graphicspath{ {Figures/} }
\usepackage{epstopdf}
\usepackage[caption=false]{subfig}
\usepackage{amsmath}
\usepackage{amsfonts}
\usepackage{amssymb}
\usepackage{bm}
\usepackage{hyperref}
\usepackage{amsthm}
\usepackage{color}

\usepackage{mathrsfs}
\usepackage[vlined,ruled]{algorithm2e}
\usepackage{url}
\usepackage{color}
\usepackage{algorithmic}
\usepackage{bbm}
\usepackage{booktabs}
\usepackage{array}
\usepackage[table]{xcolor}
\usepackage{yfonts}

\newcommand{\real}{\mathbb{R}}

\def\D{\mathrm{d}}

\begin{document}
\bibliographystyle{naturemag}

\title{Design of Large Sequential Conformational Change in Mechanical Networks}
\author{Jason Z. Kim}
\affiliation{Department of Bioengineering, University of Pennsylvania, Philadelphia, PA, 19104}
\author{Zhixin Lu}
\affiliation{Department of Bioengineering, University of Pennsylvania, Philadelphia, PA, 19104}
\author{Danielle S. Bassett}
\affiliation{Department of Bioengineering, University of Pennsylvania, Philadelphia, PA, 19104}
\affiliation{Department of Physics \& Astronomy, University of Pennsylvania, Philadelphia, PA, 19104}
\affiliation{Department of Electrical \& Systems Engineering, University of Pennsylvania, Philadelphia, PA, 19104}
\affiliation{To whom correspondence should be addressed: dsb@seas.upenn.edu}
\date{\today}
~\\
~\\
\begin{abstract}
From the complex motions of robots to the oxygen binding of hemoglobin, the function of many mechanical systems depends on large, coordinated movements of their components. Such movements arise from a network of physical interactions in the form of links that transmit forces between constituent elements. However, the principled design of specific movements is made difficult by the number and nonlinearity of interactions. Here, we model mechanical systems as linkages of rigid bonds (edges) connected by joints (nodes), and formulate a simple but powerful framework for designing full nonlinear coordinated motions using concepts from dynamical systems theory. We begin with principles for designing finite and infinitesimal motions in small modules, and show that each module is a one-dimensional map between distances across pairs of nodes. Next, we represent the act of combining modules as an iteration of this map, and design networks whose geometries reflect the map's fixed points, limit cycles, and chaos. We use this representation to design different folding sequences from a deployable network and a soliton, to a branched network acting as a mechanical AND gate. Finally, we design large changes in curvature of the entire network, and construct physical networks from laser-cut acrylic, origami, and 3D printed material to demonstrate the framework's potential and versatility for designing the full conformational trajectory of morphing metamaterials and structures.
\end{abstract}

\maketitle

\section{Introduction}
From the Mantis shrimp strike \cite{Patek2007Shrimp} and cell membrane channels \cite{Li2015Channel}, to medical stents \cite{Mori2005Stent} and solar sails \cite{Fu2016Sail}, mechanical systems are prevalent in the natural \cite{Macol2001Allostery,Burrows2013Gears} and engineered \cite{Zigoneanu2014Cloak,Surjadi2019Metamaterials} world. What makes these systems useful is their ability to change their geometry in a coordinated way to amplify motion, release pressure in cells, pass through narrow blood vessels, or dramatically increase surface area. Despite their differences, each of these systems can be commonly represented as a mechanical network, where the rigid edges encode constraints due to physical limbs or forces, and the nodes represent joints or constituent elements. A simple and powerful framework for understanding the relationship between network structure and coordinated motion is structural rigidity theory \cite{Crapo1979Rigidity}, originating from early and seminal work by J. C. Maxwell \cite{Maxwell1864Rigidity,Calladine1978Tensegrity,Jacobs1995aPebble}. Here, the number of coordinated motions is elegantly given by the difference between the numbers of node coordinates and edges. 

However, the successful design of coordinated motions depends not only on the existence of a motion, but also on the time-evolving network geometry for the duration of the motion. This changing geometry is determined by the set of node coordinates that satisfy edge constraints, just as the distance between a robot's joints are set by the length of a connecting limb. Several works provide design principles relating edge placement to node motions in small networks \cite{Kempe1877Line,Whiteley1984Bipartite,Hartenberg1964Kinematic,Kim2019Conformation}, and to detailed single-node trajectories or local perturbations in large networks \cite{Kempe1875Curves,Flechsig2017Allostery}. Other works explore lattices in the study of topological mechanics \cite{Kane2013Boundary,Mao2018Topological,Sato2018Soliton} in origami \cite{Liu2018Origami,Chen2016Origami}, along with sequential and branched motions \cite{Rafsanjani2019Shells,Coulais2017Reciprocity,Lubbers2018Branched,Stern2018Pathways,Pellegrino2001Deployable}. Excitingly, many experimental techniques are being concurrently developed to physically construct desired network geometries \cite{Coulais2018Multi,Overvelde2016Origami, Cui2019Printing,Zhao2018Printing}. With a wide range of interdisciplinary interest, it is now timely to develop a general framework for designing specific geometric trajectories in large networks.

Here, we develop such a framework by relating the geometry of a network to the progression of an iterated map in dynamical systems theory. We first draw on previous work that allows us to design the positions and velocities of the nodes in a network module along a coordinated motion \cite{Kim2019Conformation}. Then, we demonstrate that each module acts as a one-dimensional map between pairwise node distances at every point along this motion \cite{Zhou2017Maps}, and combine modules to iteratively apply this map. We tie the map's fixed points and limit cycles to crystalline states of repeating module geometries, and the stability of these points and cycles to the localization of motion to the edge or the bulk of the network. Finally, we design large changes in the shape of the entire network, and implement our framework by physically constructing networks. Hence, we design a rich and complex set of large sequential motions in networks through the dynamical properties of the map induced by a single module.

\section{Designing Coordinated Motions of a Single Module}
\begin{figure}[h!]
	\centering
	\includegraphics[width=140mm]{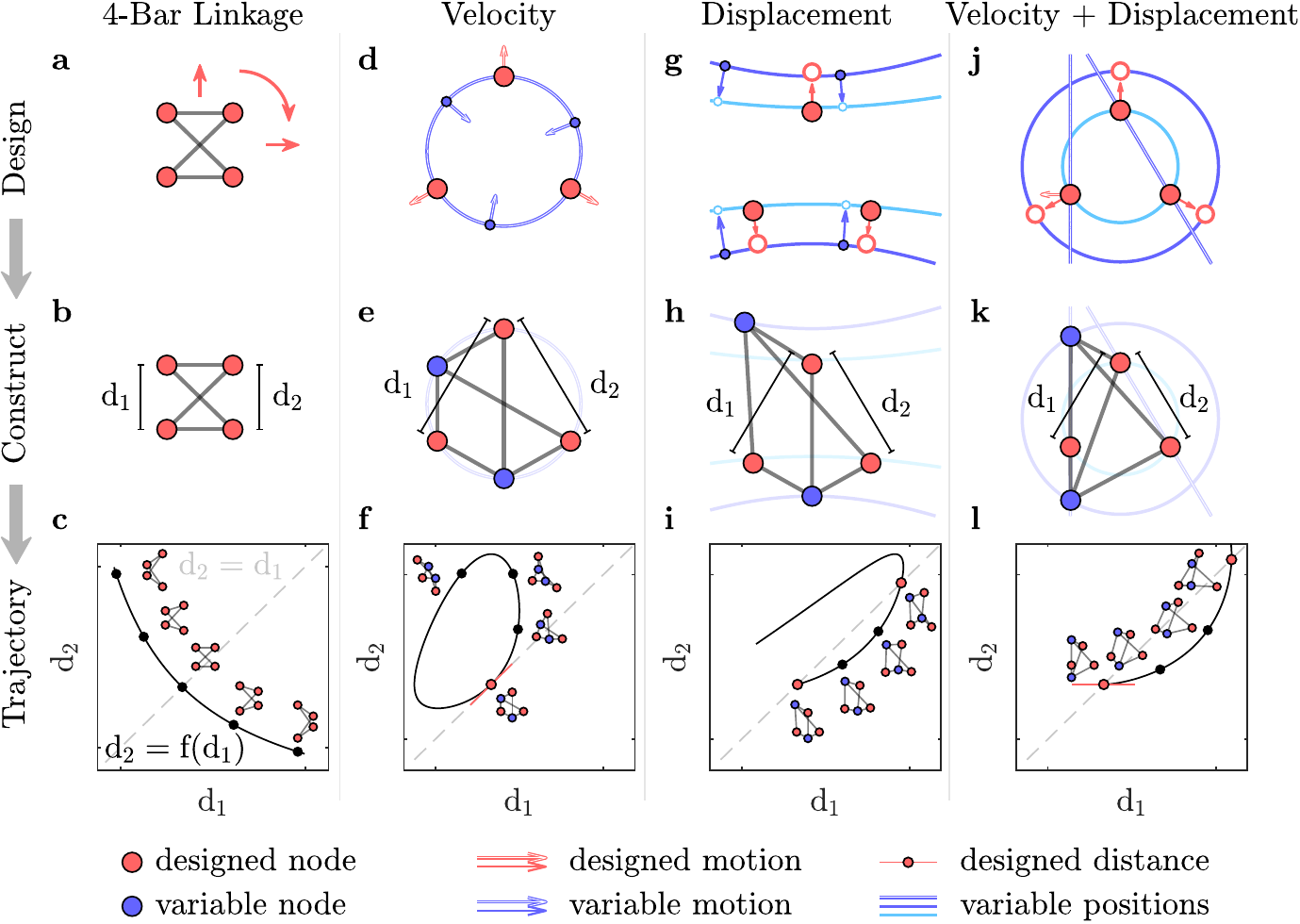}
	\caption{\textbf{Designing node velocities and displacements as a conformational motion.} (\textbf{a}) Schematic of a 4-bar linkage with $N=4$ nodes, $E=4$ edges, and $M = 4$ zero modes, with the 3 rigid-body motions indicated by red arrows. (\textbf{b}) Definition of distances $d_1$ and $d_2$ between unconnected nodes that (\textbf{c}) are plotted at each point in time $(d_1(t),d_2(t))$ as the node positions change along the conformational motion. (\textbf{d}) To design a motion, we choose the positions and velocities of designed nodes (red nodes, arrows), and solve for the positions and velocities of a fully connected variable node (blue curve, arrows) that do not change any edge lengths. (\textbf{e}) Definition of two distances $d_1$ and $d_2$ between designed nodes, with (\textbf{f}) $d_1$ and $d_2$ at the initial node positions as the red point, and the node velocities as the slope at that point. (\textbf{g}) Choice of initial (solid red) and final (hollow red) designed node positions, with solutions to the initial (dark blue) and final (light blue) variable node positions. (\textbf{h}) Constructed network where the distances $d_1$ and $d_2$ at the initial and final node positions (\textbf{i}) are shown as red dots. (\textbf{j}) Choice of designed node velocities and final positions, with corresponding solution spaces. (\textbf{k}) Placing variable nodes at the intersection of these spaces fixes the initial position, final position, and velocities of the designed nodes along the conformational motion, corresponding to (\textbf{l}) the initial point, final point, and slope of the map.}
	\label{fig:f1}
\end{figure}

To design motions in large networks, we first study the relationship between node motions and edge placement in simple network modules. As an example, consider a 4-bar linkage with $N=4$ nodes and $E=4$ edges in $d=2$ dimensions (Fig.~\ref{fig:f1}a). Each node has 2 coordinate variables $(x,y)$, and each edge adds a distance constraint between node coordinates. With $dN = 8$ variables and $E=4$ constraints, we have a $dN-E = 4$ dimensional space of allowed node coordinates. Three dimensions are the rigid body translations and rotation that exist for all 2-dimensional objects, and preserve the distances between all nodes (Fig.~\ref{fig:f1}a). The fourth defines a \emph{conformational motion} that changes distances $d_1(t)$ and $d_2(t)$ between unconnected nodes over time (Fig.~\ref{fig:f1}b). Along this motion, we plot $d_2$ against $d_1$, generating a 1-dimensional curve (Fig.~\ref{fig:f1}c) that is a \emph{map} $f$ from distance $d_1$ to distance $d_2$ at any time
\begin{align*}
d_2 = f(d_1).
\end{align*}

In a general network of $N$ nodes and $E$ edges in $d$-dimensions, the coordinates of node $i$ at time $t \geq 0$ are a vector $\bm{x}_i(t) \in \real^d$. Each rigid edge between nodes $i$ and $j$ has constant length $l_{ij}$, and adds a distance constraint on the node coordinates
\begin{align*}
l_{ij} = \|\bm{x}_i(t)-\bm{x}_j(t)\|_2.
\end{align*}
Then, the number of coordinated node motions $M$ (also called \emph{zero modes}) satisfying all edge constraints is given by generalized Maxwell counting \cite{Maxwell1864Rigidity,Calladine1978Tensegrity} as the difference between the number of coordinates $dN$ (variables) and the number of edges $E$ (constraints)
\begin{align}
\label{eq:constraint}
M = dN - E + S,
\end{align}
where $S$ is the number of states of self-stress. In our study, $S=0$ unless otherwise stated. 

From prior work, we can construct modules where we choose the positions and velocity of a set of designed nodes, fixing both the distances $d_1$ and $d_2$ and the change in distances $\delta d_1$ and $\delta d_2$ between these nodes \cite{Kim2019Conformation}. We first choose a desired initial position and velocity of the designed nodes (Fig.~\ref{fig:f1}d, red). Next, we solve for the \emph{solution space} of all positions and velocities of a fully connected variable node that together preserve the edge lengths (Fig.~\ref{fig:f1}d, blue). Finally, we add variable nodes and edges along this space until our module has 1 conformational motion (Fig.~\ref{fig:f1}e). The initial node positions fix a point $(d_1,d_2)$ on the map, and the node velocities fix the slope $\delta d_2 / \delta d_1$ of the map at this point (Fig.~\ref{fig:f1}f, red).

We can use the same method to construct modules where we choose the initial and final positions of the designed nodes at $t=0$ and $t=T$ (Fig.~\ref{fig:f1}g,h), thereby fixing the initial and final distances between the designed nodes as points $(d_1(0),d_2(0))$ and $(d_1(T),d_2(T))$ along the map (Fig.~\ref{fig:f1}i, red). We can also choose both the initial and final designed node positions, along with the node velocities, to generate two solution spaces (Fig.~\ref{fig:f1}j). By placing variable nodes at the intersection of these spaces, we fix the initial distances $(d_1(0),d_2(0))$, final distances $(d_1(T),d_2(T))$, and slope $\delta d_2 / \delta d_1$ of the map (see supplement), providing considerable design power over the shape of the map (Fig.~\ref{fig:f1}k,l, red).

\section{Module Combinations as Iterated 1-Dimensional Maps}
\begin{figure}[h!]
	\centering
	\includegraphics[width=\columnwidth]{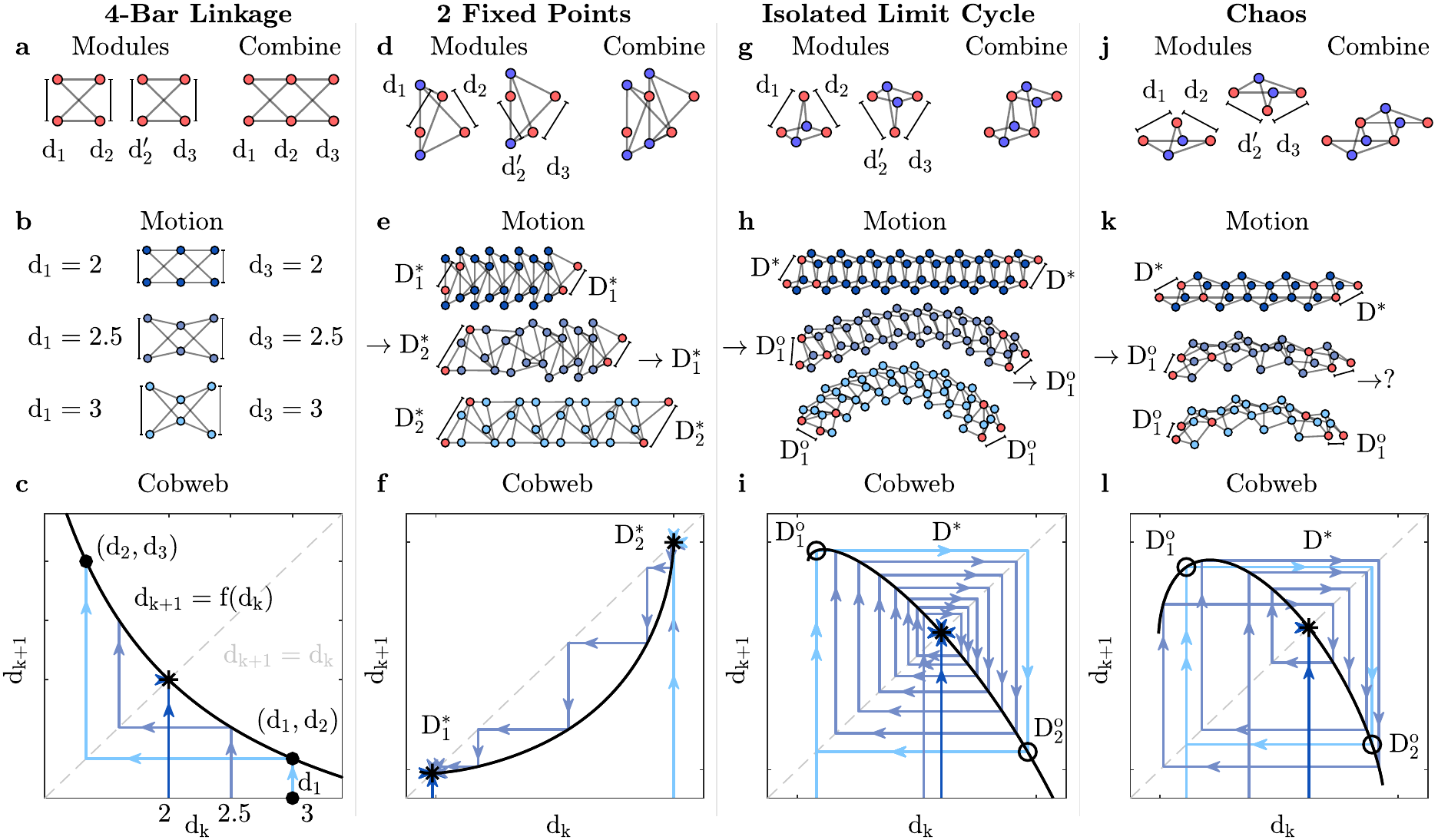}
	\caption{\textbf{Representing networks of combined modules as iterated maps.} (\textbf{a}) Two 4-bar linkage modules, combined by joining the nodes defining $d_2$ of the left module with those defining $d_2'$ of the right module, such that $d_3 = f(d_2') = f(d_2) = f(f(d_1))$. (\textbf{b}) Three network geometries at different initial distances $d_1 = 2, 2.5$, and $3$. (\textbf{c}) Cobweb plots where each set of colored arrows represents a correspondingly colored network geometry at different $d_1$. Each vertical arrow points from $d_k$ to $d_{k+1}$. (\textbf{d}) Two modules from Fig.~\ref{fig:f1}j where $d_1 = d_2$ at both initial $D_1^*$ and final $D_2^*$ geometries, that combine by joining the nodes defining $d_2$ and $d_2'$. (\textbf{e}) Network of eight modules starting at different $d_1$, with (\textbf{f}) corresponding cobweb plots showing the 8 map iterations starting at the stable fixed point $d_1 = D_1^*$ (dark blue), the unstable fixed point $d_1 = D_2^*$ (light blue), and an intermediary distance $D_1^* < d_1 < D_2^*$ (blue) where $d_k$ tends towards the stable $D_1^*$ as $k$ increases. (\textbf{g}) Alternate module combined by joining the nodes defining $d_2$ and $d_2'$. (\textbf{h}) Network of 16 modules starting at 3 different initial distances. (\textbf{i}) Cobweb plots showing 16 map iterations with initial distance at an unstable fixed point $d_1 = D^*$ (dark blue), a transition $D_o < d_1 < D^*$ (blue), and a stable period-2 cycle $d_1 = D^o_1$ (light blue). (\textbf{j}) Chaotic modules that are combined in (\textbf{k}) to form a network starting at an unstable fixed point $d_1 = D^*$ (dark blue), an unstable limit cycle $d_1 = D^o$ (light blue), and in between $D^o < d_1 < D^*$ (blue) to generate (\textbf{l}) a chaotic trajectory.}
	\label{fig:f2}
\end{figure}

Although the motion of a single module appears deceptively simple, we can design a wide range of exotic motions by defining simple rules for combining modules. Recall that in our 4-bar linkage (Fig.~\ref{fig:f1}a), we can relate the distance $d_1$ to $d_2$ with 1 application of our map $d_2 = f(d_1)$. For an identical second module with distances $d_2'$ and $d_3$ related by $d_3 = f(d_2')$, we can join these modules by combining the nodes defining $d_2'$ and $d_2$ such that $d_2' = d_2$. Then, we can relate the distance $d_3$ to $d_1$ as 2 applications of our map (Fig.~\ref{fig:f2}a)
\begin{align*}
d_3 = f(d_2') = f(d_2) = f(f(d_1)).
\end{align*}
With the $k$-th module having distances $d_k'$ and $d_{k+1}$, by joining the nodes defining $d_k'$ of the module and $d_k$ of the network, we can relate $d_1$ to $d_{k+1}$ as $k$ applications of our map
\begin{align}
\label{eq:map}
d_{k+1} = f(d_k) = \dotsm = f^k(d_1). 
\end{align}
Hence, the pairwise node distances of our combined network $(d_1,d_2,\dotsm,d_{k+1})$ is equivalent to the $k$-step trajectory of our iterated map from a specific initial distance $(d_1,f(d_1),\dotsm,f^k(d_1))$. For our combined 4-bar linkage, we consider three different geometries at initial distances $d_1 = 2$ (dark blue), $d_1 = 2.5$ (blue), and $d_1=3$ (light blue) (Fig.~\ref{fig:f2}b). For each geometry, we show the 2-step trajectory of the iterated map as arrows of the same color in a \emph{cobweb plot} \cite{Strogatz2018Nonlinear}. For module $k=1$, the arrows begin at initial distance $d_k = d_1 = 2,2.5,3$, and move up to $d_{k+1} = d_2 = f(d_1)$ to show one map iteration. For module $k=2$, the arrows move horizontally $d_k = d_2$, and up to $d_{k+1} = d_3 = f(d_2)$ as another map iteration (Fig.~\ref{fig:f2}c).

We begin with the concept of a \emph{fixed point}, defined by a distance $D^*$ that maps to itself
\begin{align}
\label{eq:fixed_point}
D^* = f(D^*).
\end{align}
At $D^*$, the network is in a \emph{crystalline state}, where the geometry of a set of modules repeats. To demonstrate, we consider the module designed in Fig.~\ref{fig:f1}j--l, where $d_1 = d_2 = D^*_1$ in the initial geometry, and $d_1 = d_2 = D^*_2$ in the final geometry. As before, we combine two modules by joining the nodes defining $d_2'$ and $d_2$ to form a network chain (Fig.~\ref{fig:f2}d). In combining eight modules by joining the nodes defining $d_i'$ and $d_i$, we form a chain with 1 conformational motion from the $D_1^*$ crystalline state (Fig.~\ref{fig:f2}e, dark-blue) to an intermediary non-crystalline state (Fig.~\ref{fig:f2}e, blue), to the $D_2^*$ crystalline state (Fig.~\ref{fig:f2}e, light-blue). In the intermediary state, the distance $d_{k+1}$ of each consecutive module $k$ moves away from $D_2^*$ and towards $D_1^*$ based on the \emph{stability} of $D_1^*$ and $D_2^*$, defined by the slope at each point
\begin{align}
\label{eq:fixed_point_stability}
s = f'(d)|_{d=D^*}.
\end{align}
If $|s| < 1$, then $D^*$ is stable, and the distance $d_{k+1}$ of consecutive modules tends toward $D^*$. If $|s| > 1$, then $D^*$ is unstable, and $d_{k+1}$ of consecutive modules moves away from $D^*$. In this example, $D_1^*$ is stable and $D_2^*$ is unstable, as seen in the cobweb plot (Fig.~\ref{fig:f2}f).

We can also design modules with period-$m$ cycles, defined by distances $D^o_1, D^o_2, \dotsm, D^o_m$ that repeat periodically every $m$ iterations
\begin{align}
\label{eq:limit_cycle}
D^o_i = f^m(D^o_i), \hspace{1cm} i = 1,\dotsm,m.
\end{align}
We show a period-2 cycle with another module (Fig.~\ref{fig:f2}g), and combine 16 modules with 1 motion that begins at $d_1=D^*$ (Fig.~\ref{fig:f2}h, dark-blue), but has another crystalline state at $d_1 = D^o_1$ where the geometry of every 2 modules repeats (Fig.~\ref{fig:f2}h, light-blue). By the chain rule, the stability of a period-$m$ cycle is the product of slopes at every point on the cycle \cite{Strogatz2018Nonlinear}
\begin{align}
\label{eq:limit_cycle_stability}
s = \prod_{i=1}^m f'(d)|_{d=D_i^o},
\end{align}
and is stable for $|s|<1$ and unstable for $|s|>1$. Here, the fixed point is unstable and the limit cycle is stable, such that consecutive modules of the intermediate network tend towards the limit cycle (Fig.~\ref{fig:f2}h,i, blue). If both the fixed point and limit cycle are unstable (Fig.~\ref{fig:f2}j--k), a network with distance $d_1$ at these points has a crystalline structure (Fig.~\ref{fig:f2}l, light and dark blue), but other distances $d_1$ yield chaotic iterative behavior with network geometries that depend sensitively on $d_1$ (Fig.~\ref{fig:f2}l, blue) with a Lyapunov exponent of $\approx$ 0.312 (see supplement). By choosing the points and slopes in the map of a single module, we design the full nonlinear motion of large networks using the behavior of the iterated map.

\section{Design of Folding Sequence}
Many recent applications such as morphing aircraft wings \cite{Sofla2010Morphing} and deployable satellite antennas \cite{Puig2010Deployable} require control over both the sequence of geometric change and the rigidity of the bulk structure. Using the dynamical principles of the previous section, we design the folding sequence of a network composed of modules by changing the stability of the module's map. At a crystalline state $d_1 = \dotsm = d_{k+1} = D^*$, we can write the change in $d_{k+1}$ with respect to $d_1$ by taking the derivative of our map Eq.~\ref{eq:map} using the chain rule
\begin{align}
\label{eq:differential}
\frac{\D}{\D d_1} d_{k+1} = \frac{\D}{\D d_1} f^k(d_1) = \prod_{i=1}^k f'(d)_{d=d_i=D^*} = s^k.
\end{align}

\begin{figure}[h!]
	\centering
	\includegraphics[width=\columnwidth]{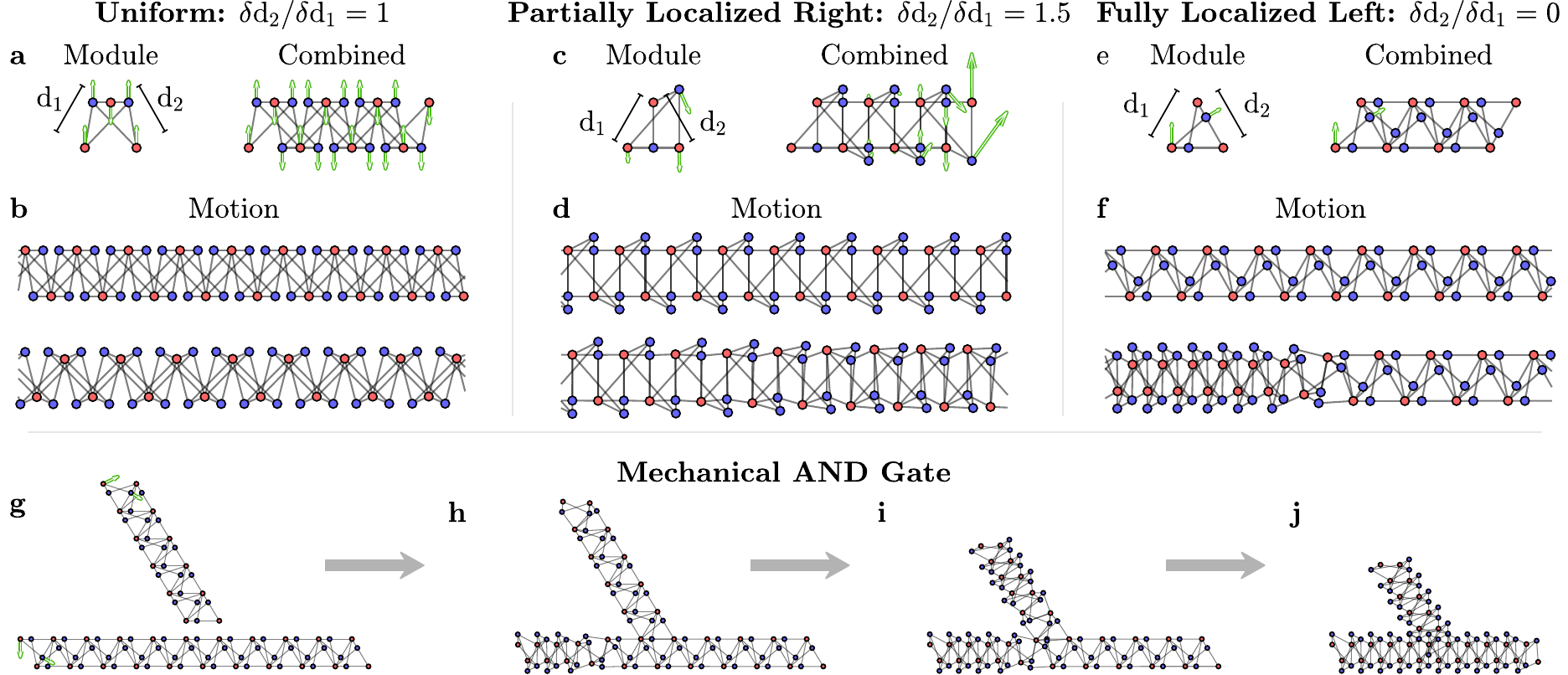}
	\caption{\textbf{Spatial localization of folding sequence determined by fixed point stability.} (\textbf{a}) Module and combined network with conformational motion designed to have velocity (green arrows) where $d_1$ and $d_2$ change equally such that the map $d_{k+1} = f(d_k)$ has a slope $s = 1$, and (\textbf{b}) motion propagates uniformly across the network. (\textbf{c}) Module and combined network where $d_2$ changes faster than $d_1$ such that the map has slope $s = 1.5$, and (\textbf{d}) the network begins deforming from the right. (\textbf{e}) Module and combined network where $d_2$ stays constant with infinitesimal change in $d_1$ such that the map has slope $s = 0$ and is \emph{super-stable}, completely isolating motion to the left. This module has two fixed points $D_1^*$ (initial) and $D_2^*$ (final), such that the combined network (\textbf{f}) collapses to the $D_2^*$ geometry from left to right. (\textbf{g}) We take two of these networks, and attach the $d_2$ end of one to the halfway point of the other, such that two infinitesimal motions exist (green arrows). These motions are mathematically coupled for finite motions, but super-stability allows (\textbf{h}) one branch to be deformed independently from the other to numerical precision, yet (\textbf{i}) both must collapse to (\textbf{j}) propagate the motion beyond the branch as a mechanical AND gate.}
	\label{fig:f3}
\end{figure}

For a system where $d_1$ and $d_2$ change identically such that $s = 1$, a unit change in $d_1$ causes a unit change in $d_{k+1}$ because $s^k = 1$, giving a uniform motion throughout the network (Fig.~\ref{fig:f3}a,b). If we increase the slope to $s = 1.5$, a unit change in $d_1$ causes a much larger change in $d_{k+1}$, localizing the majority of the motion to the $d_{k+1}$ (right) end (Fig.~\ref{fig:f3}c,d). For \emph{super-stability} where $s = 0$, any infinitesimal change in $d_1$ causes no change in $d_{k+1}$, thereby completely localizing the motion to the $d_1$ (left) end (Fig.~\ref{fig:f3}e). We can also extend sequential motion to finite deformations by using multiple fixed points. The module in Fig.~\ref{fig:f3}e has been designed to have a stable fixed point $D_1^*$, and an unstable fixed point $D_2^*$, such that the combined network collapses to this second crystalline state from $d_1$ to $d_{k+1}$ (left to right), creating a soliton that is a $D_2^*$ crystal to the left, a $D_1^*$ crystal to the right, with a transition in between that repeats with the collapse of each module \cite{Zhou2017Maps} (Fig.~\ref{fig:f3}f, \ref{fig:f5}b).

Finally, we combine these sequential chains to create branched networks that act as mechanical AND gates. We take the $d_{k+1}$-end nodes of one network from Fig.~\ref{fig:f3}e, and combine them with the middle nodes of another (Fig.~\ref{fig:f3}g) such that their floppy ends face outward (left and up). Because an \emph{infinitesimal} change in $d_1$ does not change $d_2$, we generate self-stress in Eq.~\ref{eq:constraint} and have 2 conformational motions. If we \emph{finitely} change $d_1$ at one branch, we also change the subsequent distances $d_{k+1}$, such that the motion of both branches is theoretically coupled. However, because $D_1^*$ is \emph{super-stable}, this motion does not cause a measurable change at the coupled nodes (to 64-bit precision) until one branch is almost completely collapsed (Fig.~\ref{fig:f3}h), after which we must collapse the second branch (Fig.~\ref{fig:f3}i) to collapse the whole network (Fig.~\ref{fig:f3}j). Hence, we can generate effectively independent conformational motions in branches that must all collapse for the motion to continue propagating.

\section{Design of Deployable Large-Scale Structure}
We now design the folding sequence and final geometry of combined modules to construct networks with a desired macroscopic final structure using a single actuator. Specifically, we design the curvature of a network chain's final configuration by using different modules that expand or contract the chain on either side. 

\begin{figure}[h!]
	\centering
	\includegraphics[width=1.0\columnwidth]{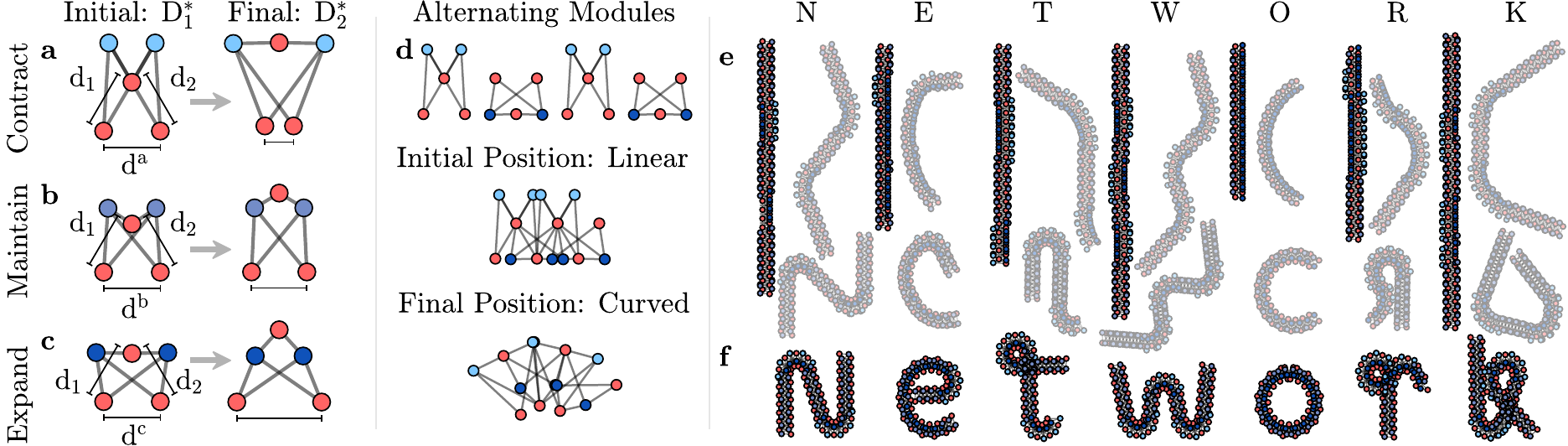}
	\caption{\textbf{Designing macroscopic network geometry through curvature.} (\textbf{a}--\textbf{c}) Network modules that transition from the same initial $D_1^*$ to final $D_2^*$ fixed points symmetrically ($d_1 = d_2$) and monotonically ($\dot{d}_1, \dot{d}_2 > 0$ from $D_1^*$ to $D_2^*$) along the conformational motion. The distance between bottom nodes either (\textbf{a}) decreases (marked here as $d_a$), (\textbf{b}) does not change (marked as $d_b$), or (\textbf{c}) increases (marked as $d_c$). (\textbf{d}) By combining modules in an alternating pattern with light-blue modules contracting the bottom nodes, and dark-blue modules expanding the top nodes, the combined network forms a line in the initial conformation, and curves downward in the final configuration. (\textbf{e}) Combined networks in their initial, intermediary, and final geometries designed to (\textbf{f}) spell out the letters in the word ``NETWORK."}
	\label{fig:f4}
\end{figure}

In three modules (Fig.~\ref{fig:f4}a--c), the initial distances $d_1 = d_2 = D_1^*$ and final distances $d_1 = d_2 = D_2^*$ are fixed points that are preserved across all modules. Further, all three modules transition monotonically such that all final conformations can be reached by increasing either $d_1$ or $d_2$ from the initial conformation, and symmetrically such that $d_1 = d_2$ throughout the motion \cite{Pellegrino2001Deployable}. Hence, the full chain can reach the final conformation by only increasing $d_1$. In the first (second, third) module (Fig.~\ref{fig:f4}a--c), the distance $d_a$ ($d_b, d_c$) between the bottom nodes decreases (does not change, increases). By combining modules in an alternating pattern, we can create portions of a network that are straight in the initial conformation, but curve in the final conformation (Fig.~\ref{fig:f4}d). As a demonstration of design capability, we create seven chains that, in their final conformation, spell out the word ``NETWORK'' (Fig.~\ref{fig:f4}e--f).

\section{Constructing Physical Networks}
\begin{figure}[h!]
	\centering
	\includegraphics[width=1.0\columnwidth]{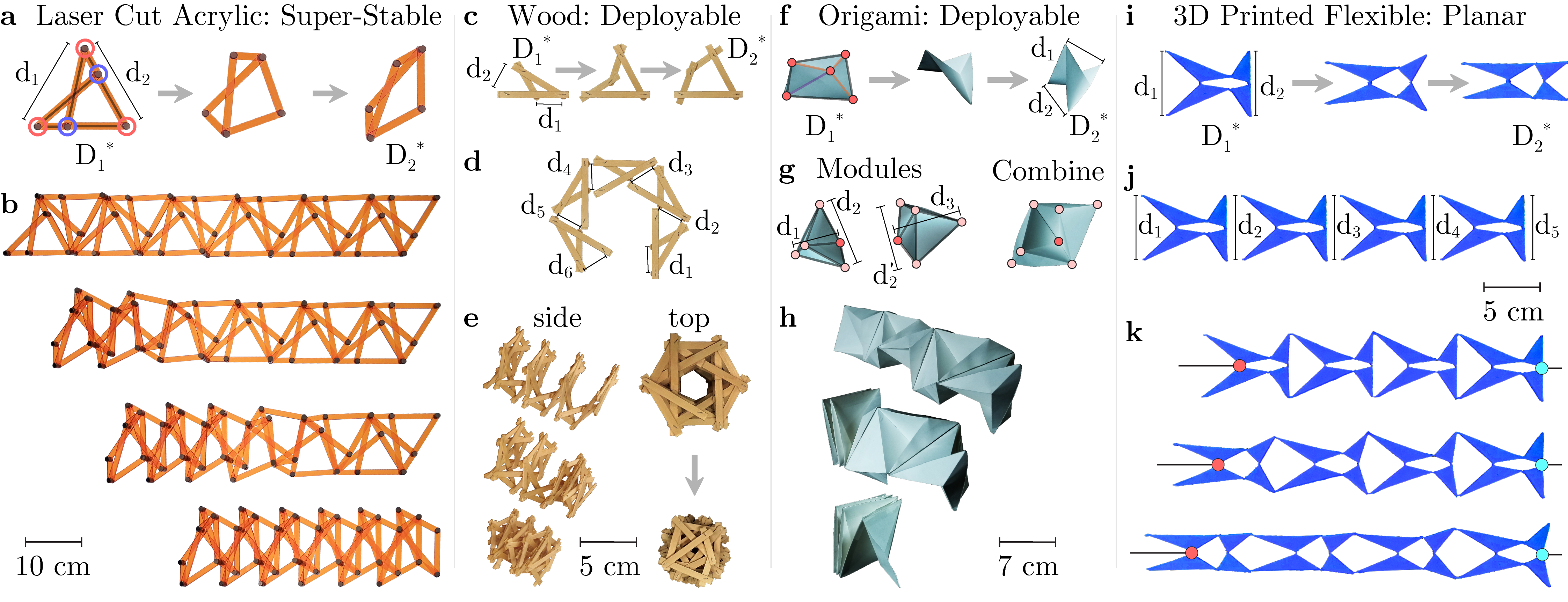}
	\caption{\textbf{Physical construction of networks.} (\textbf{a}) Photo of a super-stable module from Fig.~\ref{fig:f3}e constructed from laser-cut acrylic bars held together by Chicago screws at the joints, transitioning between two fixed points $D_1^*$ and $D_2^*$. (\textbf{b}) Photo of combined network from Fig.~\ref{fig:f3}f collapsing from $D_1^*$ to $D_2^*$. (\textbf{c}) A 4-bar linkage with two crystal states $D_1^*$ and $D_2^*$, (\textbf{d}) combined hexagonally into (\textbf{e}) an initially wide spiral helix with a channel $D_1^*$, collapsing sequentially to a narrow closed helix. (\textbf{f}) Photo of a creased square sheet of paper modeled as a linkage with 1 conformational motion moving between two crystal states $D_1^*$ and $D_2^*$ (with a mountain fold at the purple edge, and valley folds at the orange edges). (\textbf{g}) Two creased sheets combined by joining the nodes defining $d_2$ and $d_2'$, along with a third node in each module marked in bright red. (\textbf{h}) A combined network of 12 sheets that sequentially collapses from the $D_2^*$ to the $D_1^*$ flat sheet crystal state from the zero-mode localized to the left. (\textbf{i}) A 3D-printed planar module with two fixed points $D_1^*, D_2^*$. Each module is composed of triangles connected by a thin layer of material, that (\textbf{j}) form a chain where (\textbf{k}) fixing the cyan hinge and pulling the red hinge yields a sequential transition from $D_1^*$ to $D_2^*$.}
	\label{fig:f5}
\end{figure}

Here, we implement this theory for designing the geometry of both the sequence and macroscopic structure of mechanical networks by constructing physical networks. We construct the super-stable and sequentially collapsible networks from Fig.~\ref{fig:f3}e,f by laser cutting the edges from $1/8$-inch thick acrylic, and connecting their joints using Chicago screws (Fig.~\ref{fig:f5}a,b). Additionally, many deployable applications \cite{Puig2010Deployable} require a compact initial geometry and a precise, rigid final geometry. Using wooden sticks that are joined by a staple prong at the joints, we show a 4-bar linkage with two crystal state fixed points $D_1^*$ and $D_2^*$, where the $D_1^*$ point is super-stable (Fig.~\ref{fig:f5}d). These modules can be combined in a chain (Fig.~\ref{fig:f5}e) that yields a wide spiral with an open channel in the initial state $D_1^*$, and collapses to a narrow spiral with no channel in the final state $D_2^*$ (Fig.~\ref{fig:f5}f).

To demonstrate the generalizability of our framework to 3-dimensional space, we model a creased square of paper as a linkage, where each crease is a rigid edge, and the intersection of creases is a node (Fig.~\ref{fig:f5}f). We define $d_1$ and $d_2$ to be the distances between opposing corners in this sheet that collapses from the unfolded $D_1^*$ to the folded $D_2^*$ crystalline states. If we combine these modules by joining the nodes defining $d_2$ and $d_2'$ (Fig.~\ref{fig:f5}g), then we obtain an origami structure that collapses sequentially from the left end to a flat geometry (see supplement for details).

These principles also extend to planar networks comprised of polygons (e.g. triangles) connected at vertices through a thin layer of flexible material (Fig.~\ref{fig:f5}g). We designed a module with two fixed points $D_1^*$ and $D_2^*$, where the initial point $D_1^*$ is super-stable. We can chain these modules as before to yield the same iterated map $d_{k+1} = f(d_k)$ (Fig.~\ref{fig:f5}h), such that we obtain a sequential transition from $D_1^*$ to $D_2^*$ by pulling on the network (Fig.~\ref{fig:f5}i). Importantly, because this network is printed as shown, there is no required assembly.

\section{Discussion}
Ever-arising mechanical challenges \cite{Sofla2010Morphing,Puig2010Deployable} drive the development of innovative designs \cite{Overvelde2017Prism,Wei2014Polyhedra,Cheung2013Composite,Pellegrino2001Deployable}, which in turn spark novel applications \cite{Yang2015Acoustics,Cummer2016Acoustics}. In this work, we presented a simple theory for the principled design of a rich and complex set of folding sequences and large-scale geometries through the properties of a single module. Due to the practical and ubiquitous nature of linkages, these ideas are well-positioned to provide simple solutions to complex problems in robotic grasping \cite{YuZheng2005Grasp}, deployable mechanisms \cite{Puig2010Deployable}, morphing mechanical structures \cite{Sofla2010Morphing}, and tunable metamaterials \cite{Liu2018Tunable}. By writing the large, non-linear geometric conformation of a network as the iteration of one module, we retain the richness of network motion while dramatically reducing design complexity. 

Here, we studied the fundamental behaviors of this richness that directly arise from iterated maps. Immediate extensions include designing modules with complex maps (more than 2 fixed points, negative slopes at fixed points, critical slowing, bifurcations \cite{Strogatz2018Nonlinear}), and developing principles for combining modules with different maps. The theory can also extend beyond iterated maps, where linkages follow a circular path that is not formally a function ($d_2$ is not uniquely determined by $d_1$). For ease of manufacturing, previous work on planar networks \cite{Coulais2017Reciprocity} motivates the development of a module design framework specific to these systems. Finally, given the design framework for bistable linkages with elastic bonds \cite{Kim2019Conformation}, a promising future direction lies in designing tunable vibrational modes for applications in energy harvesting \cite{Liu2016Vibration} and satellite antenna \cite{Puig2010Deployable}. Hence, this simple theory provides a versatile and unifying framework for designing large sequential conformational changes in mechanical networks.

\section{Data \& Code Availability Statement}
There is no data with mandated deposition used in the manuscript or supplement. All analysis and figures were created in MATLAB, and can be publicly accessed on GitHub at \begin{verbatim}https://github.com/jk6294/Mechanical_DS.git\end{verbatim} with a test script that will exactly replicate and save all figures in the manuscript except the construction of physical networks.

\section{Acknowledgments}
\noindent We gratefully acknowledge Melody X. Lim, Ann E. Sizemore, Lia Papadopoulos, Jennifer Stiso, Harang Ju, and Erin G. Teich for conversations and comments on the manuscript. JZK acknowledges support from the NIH T32-EB020087, PD: Felix W. Wehrli, and the National Science Foundation Graduate Research Fellowship No. DGE-1321851. DSB acknowledges support from the John D. and Catherine T. MacArthur Foundation, the ISI Foundation, the Alfred P. Sloan Foundation, an NSF CAREER award PHY-1554488, and from the NSF through the University of Pennsylvania Materials Research Science and Engineering Center (MRSEC) DMR-1720530.

\newpage
\section{References}
\bibliography{references}

\end{document}


\bibliographystyle{naturemag}

\title{Supplementary Information: Design of Large Sequential Conformational Change in Mechanical Networks}
\author{Jason Z. Kim}
\affiliation{Department of Bioengineering, University of Pennsylvania, Philadelphia, PA, 19104}
\author{Zhixin Lu}
\affiliation{Department of Bioengineering, University of Pennsylvania, Philadelphia, PA, 19104}
\author{Danielle S. Bassett}
\affiliation{Department of Bioengineering, University of Pennsylvania, Philadelphia, PA, 19104}
\affiliation{Department of Physics \& Astronomy, University of Pennsylvania, Philadelphia, PA, 19104}
\affiliation{Department of Electrical \& Systems Engineering, University of Pennsylvania, Philadelphia, PA, 19104}
\affiliation{To whom correspondence should be addressed: dsb@seas.upenn.edu}
\date{\today}
~\\
~\\

\maketitle
\tableofcontents
\newpage

\section{General Linkage Framework}
The design and motion of linkages can be thought of as solutions to variables subject to distance constraints. For a system of $N$ nodes $\mc{V} = \{1,\dotsm,N\}$ in $d$-dimensions where the $i$-th node has coordinate $\bm{x}_i \in \real^d$, the variables are the node coordinates $\bm{x} \in \real^{dN}$
\begin{align*}
\bm{x} = 
\begin{bmatrix}
\bm{x}_1\\
\bm{x}_2\\
\vdots\\
\bm{x}_N
\end{bmatrix}.
\end{align*}
If we connect node pairs with a set of $E$ edges $\mc E \subseteq \mc V \times \mc V$, then an edge $k$ between nodes $i$ and $j$ has squared length $l_k^2 = (\bm{x}_i-\bm{x}_j)^{\top} (\bm{x}_i-\bm{x}_j)$. Then the constraints are the distances
\begin{align}
\label{eq:constraint}
\bm{l} = 
\begin{bmatrix}
l_1^2\\
l_2^2\\
\vdots\\
l_E^2
\end{bmatrix}
= 
\bm{f}(\bm{x}),
\end{align}
where $\bm{f}(\bm{x})$ measures the distance of each pair of connected nodes. Hence, the simplified form of Maxwell counting is a difference between variables and constraints. For $dN$ node coordinates (variables) subject to $E$ edge constraints, the dimension of solutions is generally
\begin{align}
\label{eq:maxwell1}
M = dN - E,
\end{align}
where $M$ is the dimension of independent node motions satisfying edge constraints. More formally, our configuration space consists of $dN$ node coordinates, and each edge constraint $l_k^2 - (\bm{x}_i-\bm{x}_j)^\top(\bm{x}_i-\bm{x}_j) = 0$ defines an \emph{algebraic variety} that is the set of node coordinates satisfying the edge constraint. Each constraint defines a variety that has dimension $dN-1$, and the general intersection of $E$ varieties of dimension $dN-1$ is $dN-E$. While this statement is generally true, there are pathological cases where it is not. For example, in 3-dimensions, two non-parallel planes will always intersect at a line. However, a parabaloid may only intersect a plane at a single point, thereby violating Eq.~\ref{eq:maxwell1}, and generating self-stress. Hence, we are motivated to find a more general formulation of Eq.~\ref{eq:maxwell1}. 
\newpage

\section{Motions and States of Self-Stress}
To see where this counting scheme fails, let us consider the set of infinitesimal displacements in time, $\dot{\bm{x}}$, that satisfy the distance constraints to \emph{linear} order. We take the differential with respect to time of edge length $k$, and enforcing the differential $\dot{l}_k = 0$ to linear order
\begin{equation}
\begin{aligned}
\label{eq:infinitesimal}
\frac{\delta}{\delta t} l_k^2 &= \frac{\delta}{\delta t} (\bm{x}_i-\bm{x}_j)^{\top}(\bm{x}_i-\bm{x}_j)\\
2l_k\dot{l} &= 2(\bm{x}_i-\bm{x}_j)^{\top}(\dot{\bm{x}}_i-\dot{\bm{x}}_j)\\
0 &= (\bm{x}_i-\bm{x}_j)^{\top}(\dot{\bm{x}}_i-\dot{\bm{x}}_j).
\end{aligned}
\end{equation}
At any particular configuration of nodes $\bm{x} = \bm{x}^*$, we can treat the node velocities $\dot{\bm{x}}$ as variables, and notice that the velocities are linear with respect to the positions in Eq.~\ref{eq:infinitesimal}. We can bring together these linear constraints in matrix form
\begin{align}
\label{eq:rigidity}
\bm{0} = R\dot{\bm{x}},
\end{align}
where $R = R(\bm{x} = \bm{x}^*)$ is the \emph{rigidity matrix} of size $E \times dN$ with mostly zeros, except in the $k$-th row containing $(\bm{x}_i^*-\bm{x}_j^*)^{\top}$ multiplied by $\dot{\bm{x}}_i$, and $(\bm{x}_j^*-\bm{x}_i^*)^{\top}$ multiplied by $\dot{\bm{x}}_j$. Then the \emph{nullspace} of this matrix, $\mc N(R)$, provides the set of all node velocities that satisfy the distance constraints to linear order (Eq.~\ref{eq:rigidity}). In the general case for most node positions $\bm{x}^*$, the rigidity matrix has full rank, and the number of independent node motions (given by the dimension of the nullspace) is simply
\begin{align*}
\dim(\mc N(R)) = dN - E,
\end{align*}
in accordance with Maxwell counting (Eq.~\ref{eq:maxwell1}). However, there exists a small set of pathological node positions that cause $R$ to lose rank, such that $\dim(\mc N(R)) > dN - E$. In this case, we have a \emph{state of self-stress} (SSS), where there is a greater number of infinitesimal motions than expected. This issue motivates the definition of generalized Maxwell counting, where given $S$ SSS, the number of independent motions to linear order is given by
\begin{align}
\label{eq:maxwell2}
M = dN - E + S.
\end{align}
Unfortunately, the ability of these extra motions to extend into finite deformations is complicated, and falls under the domain of higher-order rigidity and bifurcation theory. 
\newpage

\section{Elaborating on Single Module Design: Velocity}
The module design process presented in the main text arises from a slight reformulation of this linkage framework. Previously, we fixed all node positions $\bm{x}^*$, treated the node velocities as variables $\dot{\bm{x}}$, and solved for the velocities that preserve edge length to linear order through the nullspace of the rigidity matrix (Eq.~\ref{eq:rigidity}). Here, we partition the nodes into two disjoint parts, $\mc V = \mc V_S \cup \mc V_U$: the specified nodes $\mc V_S$ with corresponding positions and motions $\bm{x}_S$ and $\dot{\bm{x}}_S$, and the unspecified nodes $\mc V_U$ with corresponding positions and motions $\bm{x}_U$ and $\dot{\bm{x}}_U$. Further, we assume that connections only exist between the specified and unspecified nodes $\mc E = \mc V_S \times \mc V_U$ to form a bipartite graph.

Consider the simple case of $n$ specified nodes and 1 unspecified node. We can then write the linearized constraints as
\begin{align*}
\bm{0} = 
\begin{bmatrix}
(\bm{x}_{S1}-\bm{x}_{U1})^{\top}(\dot{\bm{x}}_{S1}-\dot{\bm{x}}_{U1})\\
(\bm{x}_{S2}-\bm{x}_{U1})^{\top}(\dot{\bm{x}}_{S2}-\dot{\bm{x}}_{U1})\\
\vdots\\
(\bm{x}_{Sn}-\bm{x}_{U1})^{\top}(\dot{\bm{x}}_{Sn}-\dot{\bm{x}}_{U1})
\end{bmatrix}.
\end{align*}
If we fix the specified node positions $\bm{x}_S = \bm{x}_S^*$ and motions $\dot{\bm{x}}_S = \dot{\bm{x}}_S^*$ as constants, then
\begin{align*}
\bm{0} = 
\begin{bmatrix}
(\bm{x}_{S1}^*-\bm{x}_{U1})^{\top}(\dot{\bm{x}}_{S1}^*-\dot{\bm{x}}_{U1})\\
(\bm{x}_{S2}^*-\bm{x}_{U1})^{\top}(\dot{\bm{x}}_{S2}^*-\dot{\bm{x}}_{U1})\\
\vdots\\
(\bm{x}_{Sn}^*-\bm{x}_{U1})^{\top}(\dot{\bm{x}}_{Sn}^*-\dot{\bm{x}}_{U1})
\end{bmatrix}.
\end{align*}
We expand each equation and pull out the variable unspecified positions and motions
\begin{align*}
\bm{0} = 
\begin{bmatrix}
\bm{x}_{S1}^{*\top}\dot{\bm{x}}_{S1}^*\\
\bm{x}_{S2}^{*\top}\dot{\bm{x}}_{S2}^*\\
\vdots\\
\bm{x}_{Sn}^{*\top}\dot{\bm{x}}_{Sn}^*
\end{bmatrix}
-
\begin{bmatrix}
\bm{x}_{S1}^{*\top}\\
\bm{x}_{S2}^{*\top}\\
\vdots\\
\bm{x}_{Sn}^{*\top}
\end{bmatrix}
\dot{\bm{x}}_{U1}
-
\begin{bmatrix}
\dot{\bm{x}}_{S1}^{*\top}\\
\dot{\bm{x}}_{S2}^{*\top}\\
\vdots\\
\dot{\bm{x}}_{Sn}^{*\top}
\end{bmatrix}
\bm{x}_{U1}
+
\bm{x}_{U1}^{\top}\dot{\bm{x}}_{U1}
\begin{bmatrix}
1\\1\\\vdots\\1
\end{bmatrix},
\end{align*}
and notice that there is only 1 nonlinear term in this system, namely $c = \bm{x}_{U1}^{\top}\dot{\bm{x}}_{U1}$. If we temporarily omit this nonlinearity, and substitute $c$ as a free variable, we can write the linearized constraint equations as a linear system of equations $\bm{b} = A\bm{v}$
\begin{align}
\label{eq:bipartite_vel_linear}
\underbrace{\begin{bmatrix}
\bm{x}_{S1}^{*\top}\dot{\bm{x}}_{S1}^*\\
\bm{x}_{S2}^{*\top}\dot{\bm{x}}_{S2}^*\\
\vdots\\
\bm{x}_{Sn}^{*\top}\dot{\bm{x}}_{Sn}^*
\end{bmatrix}}_{\bm{b}}
=
\underbrace{
\begin{bmatrix}
\dot{\bm{x}}_{S1}^{*\top} & \bm{x}_{S1}^{*\top} & -1\\
\dot{\bm{x}}_{S2}^{*\top} & \bm{x}_{S2}^{*\top} & -1\\
\vdots & \vdots & \vdots\\
\dot{\bm{x}}_{Sn}^{*\top} & \bm{x}_{Sn}^{*\top} & -1
\end{bmatrix}}_{A}
\underbrace{
\begin{bmatrix}
\bm{x}_{U1}\\
\dot{\bm{x}}_{U1}\\
c
\end{bmatrix}}_{\bm{v}}.
\end{align}
Then, the unspecified node positions and motions arise as the particular $\bm{v}_P = A^+\bm{b}$ and homogeneous $\bm{v}_H \in \mc N(A)$ solutions to Eq.~\ref{eq:bipartite_vel_linear} where
\begin{align*}
\bm{v} = \bm{v}_P + \bm{v}_H,
\end{align*}
that satisfy the one nonlinear constraint $c = \bm{x}_{U1}^{\top}\dot{\bm{x}}_{U1}$. If an unspecified node that is connected to all specified nodes is placed along this solution space, then the specified node positions $\bm{x}_S^*$ and motions $\dot{\bm{x}}_S^*$ satisfy the edge constraints. By placing enough unspecified nodes and edges along this space such that $M = 4$, the only remaining conformational motion becomes $\dot{\bm{x}}_S$. 
\newpage

\section{Elaborating on Single Module Design: Displacement}
Much in the same way, we can design the initial and final positions of the specified nodes. Consider again our bipartite graph of $n$ specified nodes with initial $\bm{x}^0_S$ and final $\bm{x}^*_S$ positions, fully connected to an unspecified node with initial $\bm{x}^0_{U1}$ and final $\bm{x}^*_{U1}$ positions. The constraint we must satisfy is that the edge lengths at the initial and final positions remain constant, such that
\begin{align*}
\begin{bmatrix}
(\bm{x}_{S1}^0-\bm{x}_{U1}^0)^{\top}(\bm{x}_{S1}^0-\bm{x}_{U1}^0)\\
(\bm{x}_{S2}^0-\bm{x}_{U1}^0)^{\top}(\bm{x}_{S2}^0-\bm{x}_{U1}^0)\\
\vdots\\
(\bm{x}_{Sn}^0-\bm{x}_{U1}^0)^{\top}(\bm{x}_{Sn}^0-\bm{x}_{U1}^0)
\end{bmatrix}
=
\begin{bmatrix}
(\bm{x}_{S1}^*-\bm{x}_{U1}^*)^{\top}(\bm{x}_{S1}^*-\bm{x}_{U1}^*)\\
(\bm{x}_{S2}^*-\bm{x}_{U1}^*)^{\top}(\bm{x}_{S2}^*-\bm{x}_{U1}^*)\\
\vdots\\
(\bm{x}_{Sn}^*-\bm{x}_{U1}^*)^{\top}(\bm{x}_{Sn}^*-\bm{x}_{U1}^*)
\end{bmatrix}.
\end{align*}
We can again expand these terms to yield
\begin{align*}
\begin{bmatrix}
\bm{x}_{S1}^{0\top}\bm{x}_{S1}^0\\
\bm{x}_{S2}^{0\top}\bm{x}_{S2}^0\\
\vdots\\
\bm{x}_{Sn}^{0\top}\bm{x}_{Sn}^0
\end{bmatrix}
-
2\begin{bmatrix}
\bm{x}_{S1}^{0\top}\\
\bm{x}_{S2}^{0\top}\\
\vdots\\
\bm{x}_{Sn}^{0\top}
\end{bmatrix}
\bm{x}_{U1}^0
+
\bm{x}_{U1}^{0\top}\bm{x}_{U1}^0
\begin{bmatrix}
1\\1\\\vdots\\1
\end{bmatrix}
=
\begin{bmatrix}
\bm{x}_{S1}^{*\top}\bm{x}_{S1}^*\\
\bm{x}_{S2}^{*\top}\bm{x}_{S2}^*\\
\vdots\\
\bm{x}_{Sn}^{*\top}\bm{x}_{Sn}^*
\end{bmatrix}
-
2\begin{bmatrix}
\bm{x}_{S1}^{*\top}\\
\bm{x}_{S2}^{*\top}\\
\vdots\\
\bm{x}_{Sn}^{*\top}
\end{bmatrix}
\bm{x}_{U1}^*
+
\bm{x}_{U1}^{*\top}\bm{x}_{U1}^*
\begin{bmatrix}
1\\1\\\vdots\\1
\end{bmatrix},
\end{align*}
and rearrange to isolate the constants $\bm{x}_S^0,\bm{x}_S^*$ from the variables $\bm{x}_U^0,\bm{x}_U^*$ to yield another linear equation, with a similarly substituted nonlinear free variable $c = \bm{x}_{U1}^{0\top}\bm{x}_{U1}^0 - \bm{x}_{U1}^{*\top}\bm{x}_{U1}^*$
\begin{align}
\label{eq:bipartite_disp_linear}
\underbrace{
\begin{bmatrix}
\bm{x}_{S1}^{0\top}\bm{x}_{S1}^0-\bm{x}_{S1}^{*\top}\bm{x}_{S1}^*\\
\bm{x}_{S2}^{0\top}\bm{x}_{S2}^0-\bm{x}_{S2}^{*\top}\bm{x}_{S2}^*\\
\vdots\\
\bm{x}_{Sn}^{0\top}\bm{x}_{Sn}^0-\bm{x}_{Sn}^{*\top}\bm{x}_{Sn}^*
\end{bmatrix}}_{\bm{b}}
=
\underbrace{
\begin{bmatrix}
2\bm{x}_{S1}^{0\top} & -2\bm{x}_{S1}^* & -1\\
2\bm{x}_{S2}^{0\top} & -2\bm{x}_{S2}^* & -1\\
\vdots\\
2\bm{x}_{Sn}^{0\top} & -2\bm{x}_{Sn}^* & -1
\end{bmatrix}}_{A}
\underbrace{
\begin{bmatrix}
\bm{x}_{U1}^0\\
\bm{x}_{U1}^*\\
c
\end{bmatrix}}_{\bm{v}}.
\end{align}
Again, we see that $\bm{b}$ and $A$ only contain specified node initial and final positions that we fix as constants, such that the equation is linear in our unspecified node initial and final positions, $\bm{v}$. Our solution space is again given by a particular and homogeneous solution $\bm{v} = \bm{v}_P + \bm{v}_H$ that satisfies one quadratic constraint $c = \bm{x}_{U1}^{0\top}\bm{x}_{U1}^0 - \bm{x}_{U1}^{*\top}\bm{x}_{U1}^*$. By placing unspecified nodes and edges on this space such that $M=4$, we have 1 conformational motion where the desired initial $\bm{x}^0$ and final $\bm{x}^*$ have the same edge lengths.
\newpage

\section{Module Combination Maintains 1 Conformational Motion}
Throughout the main text, we combine modules in 2-dimensional space, each with 1 conformational motion, by joining pairs of nodes. Here, we will demonstrate why the combined network retains 1 conformational motion. As we previously mentioned, each module of $N$ nodes and $E$ edges in $d=2$ dimensional space has $dN$ coordinate variables subject to $E$ constraints, for a total of $dN-E = 4$ dimensions of possible node motions. Three of these motions are the rigid body translations and rotation that are present in all objects in 2 dimensional space, and the fourth is the conformational motion.

In 2-dimensional space, consider two modules that each have 1 conformational motion. Then the first module has $M_1 = 4$ motions, and second module has $M_2 = 4$ motions, such that a system with both modules has $M_c = M_1 + M_2 = 8$ motions. If we combine these modules by joining two pairs of nodes, $(i_1,i_2)$ from modules 1 and 2, as well as $(j_1,j_2)$ from modules 1 and 2, then we remove 4 variables because the $x$- and $y$-coordinates of nodes $i_1$ and $i_2$ combine, and the $x$- and $y$-coordinates of nodes $j_1$ and $j_2$ combine. Hence, the combined network will have $M_c = M_1 + M_2 - 4 = 4$ motions. Three of these motions must be the rigid body translations and rotation, and the fourth is the conformational motion. 
\newpage

\section{Numerical Form of the Iterated Map}
Throughout the main text, we show the 1-dimensional curve of distances $(d_1,d_2)$ between unconnected nodes of a module as an iterated map. To numerically obtain this curve, we perform a 4-th order Runge-Kutta numerical integration of the conformational motion. To introduce some notation, we will use $\bm{x}^t$ to represent all node coordinates in a module at time $t$. Further, we will use column vectors $\bm{s}_x(\bm{x}^t),\bm{s}_y(\bm{x}^t)$ and $\bm{s}_r(\bm{x}^t)$ to represent the rigid body translational and rotational motion at $\bm{x}^t$, and collect them into a matrix $S(\bm{x}^t) = [\bm{s}_x(\bm{x}^t), \bm{s}_y(\bm{x}^t), \bm{s}_r(\bm{x}^t)]$. Finally, knowing the node coordinates at $\bm{x}^t$ is sufficient to determine the entries of the rigidity matrix $R(\bm{x}^t)$ as shown in Eq.~\ref{eq:rigidity}.

In Eq.~\ref{eq:rigidity}, we show that the node motions which preserve edge lengths to linear order are given by the nullspace of the rigidity matrix. Our modules are designed to have 4 possible motions, such that the nullspace of $R(\bm{x}^t)$ contains 4 linearly independent vectors that we make orthonormal, and collect column-wise into a matrix $W(\bm{x}^t)$. By projecting our rigid body motions onto these nullspace vectors, we obtain the coefficients of our nullspace vectors $S^\top(\bm{x}^t) W(\bm{x}^t)$ that reconstruct the rigid body motions. Then the coefficients of our nullspace vectors that reconstruct the conformational motion are orthogonal to $S^\top(\bm{x}^t) W(\bm{x}^t)$, and are given by $\bm{c}(\bm{x}^t) = \mc N(S^\top(\bm{x}^t) W(\bm{x}^t))$, such that the conformational motion at $\bm{x}^t$ is given by
\begin{align}
\label{eq:conformational_nullspace}
\dot{\bm{x}}^t = \bm{g}(\bm{x}^t) = W(\bm{x}^t)\bm{c}(\bm{x}^t).
\end{align}
Now that we have the instantaneous node velocities as a nonlinear function $\bm{g}$ of node positions at time $t$, we are able to use the RK4 algorithm to numerically integrate the node positions from the initial node positions $\bm{x}^0$. Then, given the trajectory of node coordinates $\bm{x}^t$, we compute our 1-dimensional map by solving for the distances between unconnected nodes $d_1(t)$ and $d_2(t)$ at each time point $t$.  Further, we can quantify the error of this numerical integration at any position $\bm{x}^t$ by calculating the deviation of edge lengths at time $t$ from the known edge lengths at time $t=0$. We note that this same procedure is used to simulate forward the full network of combined modules with 1 conformational motion.
\newpage

\section{Analytic Form of the Iterated Map}
While numerical integration yields points along the 1-dimensional map of conformational motions over time given by $d_1(t)$ and $d_2(t)$, an analytical form of the map is desired for numerically sensitive applications such as computing the Lyapunov exponent of a map. Specifically, we would like the equational form of $f$ for $d_2 = f(d_1)$. We solve for this equational form in our systems of 5 nodes (3 designed, 2 variable) and 6 edges. Without loss of generality, we number the designed nodes such that $d_1$ is the distance between nodes $1$ and $2$, and $d_2$ is the distance between nodes $2$ and $3$, and we number the variable nodes as 4 and 5. Further, we can always find a set of rigid body motions to place node 2 at the coordinates $(x_2,y_2) = (0,0)$, and place node 1 at the coordinates $(x_1,y_1) = (d_1,0)$.

We begin by writing the variable node positions as a function of $(x_1,y_1)$ and $(x_2,y_2)$ through the distance constraints from the edge lengths. Specifically, nodes $i=4,5$ satisfy
\begin{align*}
l_{1i}^2 &= (x_i-x_1)^2 + (y_i-y_1) = x_i^2 + y_i^2 - 2d_1x_i + d_1^2 ,\\
l_{2i}^2 &= (x_i-x_2)^2 + (y_i-y_2) = x_i^2 + y_i^2,
\end{align*}
and subtracting the two equations, we can solve for $x_i$, and resubstitute to solve for $y_i$
\begin{align*}
x_i = \frac{d_1^2 + l_{2i}^2 - l_{1i}^2}{2d_1}
\hspace{1in}
y_i = \pm\sqrt{l_{2i}^2-\left(\frac{d_1^2 + l_{2i}^2 - l_{1i}^2}{2d_1}\right)^2}.
\end{align*}
From the known initial position of the nodes, $\bm{x}^0$, we can determine whether $y_i$ is positive or negative. Then, we can write the position of node 3 as satisfying constraints
\begin{align*}
l_{34}^2 &= (x_3-x_4)^2 + (y_3-y_4)^2\\
l_{35}^2 &= (x_3-x_5)^2 + (y_3-y_5)^2,
\end{align*}
which are the equations for two circles, one centered at $(x_4,y_4)$ with radius $l_{34}$, and another centered at $(x_5,y_5)$ with radius $l_{35}$. The intersection of these circles yields $x_3$ and $y_3$, and is solved symbolically as a function of $d_1$. Because $d_2$ is the distance between nodes 2 and 3, and node 2 is located at $(0,0)$, the distance $d_2$ is simply the length of the position of node 3, finally yielding the map $d_2 = f(d_1) = \sqrt{x_3^2(d_1) + y_3^2(d_1)}$.
\newpage

\section{Condition for a Conformational Motion to Act as a Map}
\begin{figure}[h!]
	\centering
	\includegraphics[width=\columnwidth]{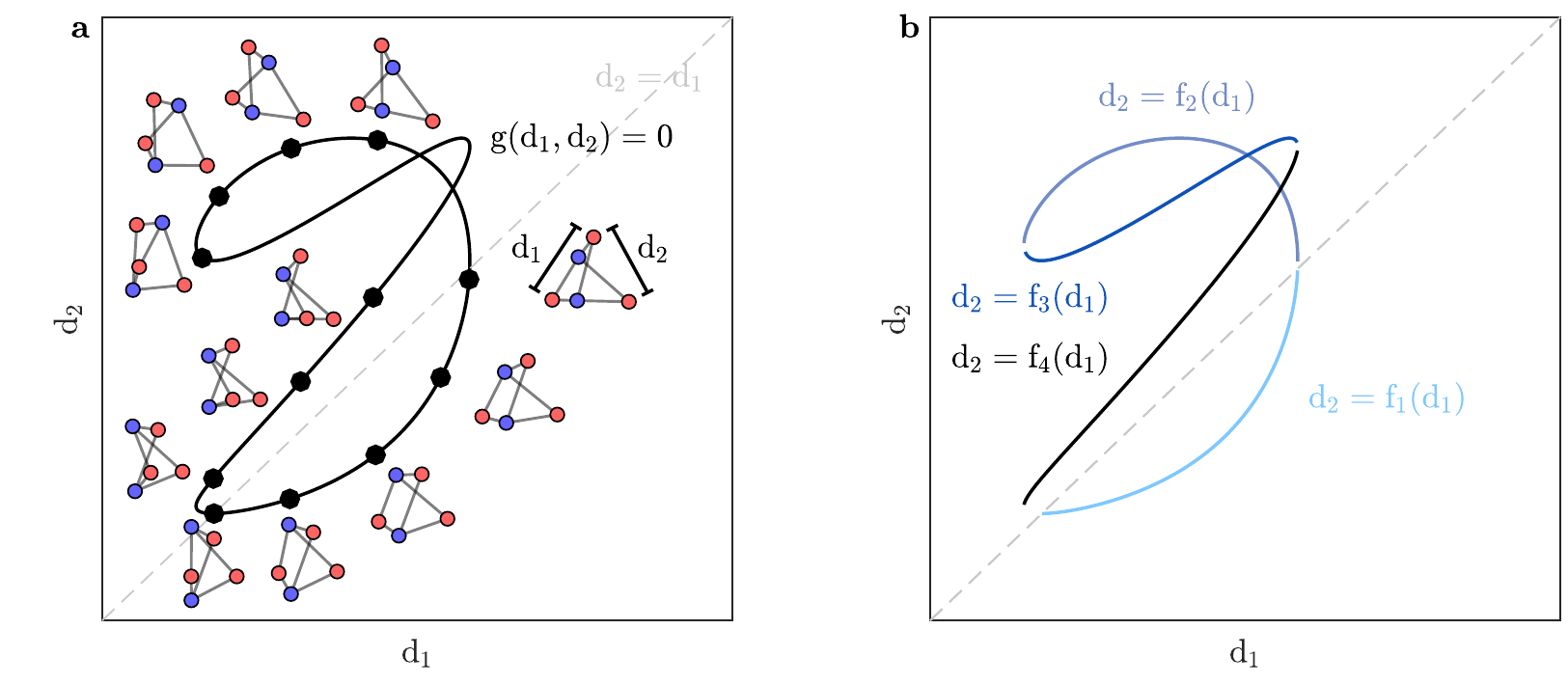}
	\caption{\textbf{Decomposition of a conformational motion into valid map segments.} (\textbf{a}) A plot of distances $d_1$ \emph{versus} $d_2$ between unconnected red nodes in a module along the full conformational motion, with example network geometries at select points. (\textbf{b}) Decomposition of the curve $(d_1,d_2)$ into four segments that uniquely map distance $d_1$ to distance $d_2$.}
	\label{fig:f1supp}
\end{figure}

Throughout the main text, we consider the implications of viewing module combinations as iterated maps. For this perspective to hold rigorously true, we must write $d_2$ as a proper function of $d_1$, where each value of $d_1$ uniquely determines a value of $d_2$. This property generally does not hold true along the full trajectory of a single module. For example, consider the following module from the main text (Fig.~\ref{fig:f1supp}a), with 1 conformational motion, and distances $d_1$ and $d_2$ between unconnected red nodes. As the node coordinates change along this conformational motion, we can plot the distance pair $(d_1,d_2)$ as a continuous 1-dimensional curve that is some function of $d_1$ and $d_2$, shown in black. Immediately, we notice that for many values of $d_1$, there are many different possible values of $d_2$ on the curve. For the iterated map perspective to hold true, we must select a continuous segment of this curve where $d_2$ is uniquely determined by $d_1$. For this particular module, we can select 4 such segments $f_1$ (light blue), $f_2$ (blue), $f_3$ (dark blue), $f_4$ (black) as 4 valid maps from $d_1$ to $d_2$ (Fig.~\ref{fig:f1supp}b). We ensure that the combined networks of the main text are valid functions.
\newpage

\section{Elaborating on Cobweb Plots}
\begin{figure}[h!]
	\centering
	\includegraphics[width=\columnwidth]{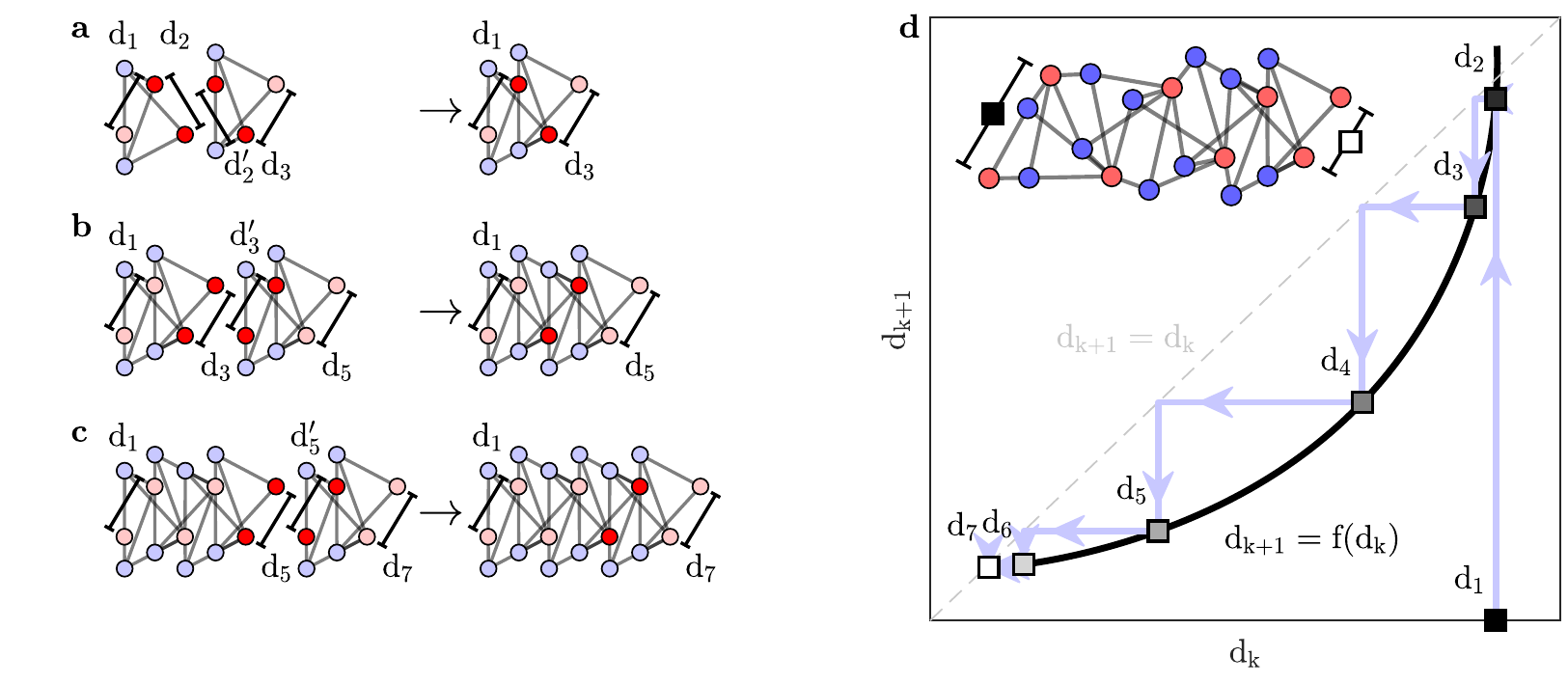}
	\caption{\textbf{Module combination and map iteration.} (\textbf{a}) The combination of two modules from Figure \ref{fig:f1supp}a, where the bright red nodes defining $d_2$ of the left module, and those defining $d_2'$ of the right module are combined. (\textbf{b}) Two combined networks from panel (\textbf{a}), where the bright red nodes defining $d_3$ and $d_3'$ are joined. (\textbf{c}) The combined networks from panels (\textbf{a}) and (\textbf{b}), where the bright red nodes defining $d_5$ and $d_5'$ are joined. (\textbf{d}) A cobweb plot of the network in panel (\textbf{c}), where the initial distance $d_1$ is marked as a black square, and the final distance $d_7$ is marked in white. The arrows begin at $d_1$, and move up to $d_2 = f(d_1)$ to a dark gray square representing $d_2$. Then, the arrows move left and down to $d_3 = f(d_2)$. Subsequent left and down arrows correspond to another iteration of the map $d_{k+1} = f(d_k)$, which are the distances between unconnected nodes in the network.}
	\label{fig:f2supp}
\end{figure}

In the main text, we discuss the representation of the conformation of a network of combined modules using a cobweb plot. Here we go into more detail with the module used in the previous section. We begin with two of these modules, where the nodes to be combined are colored in bright red, and these nodes define the distance $d_1$ in the left module, and the distance $d_2$ in the right module (Fig.~\ref{fig:f2supp}a). We can take two of these combined networks (Fig.~\ref{fig:f2supp}a) with distances $d_1$ and $d_3$ for the left network, and $d_3'$ and $d_5$ for the right module, and combine them at the bright red nodes defining $d_3$ and $d_3'$ (Fig.~\ref{fig:f2supp}b). We can extend our network by continuing to append modules (Fig.~\ref{fig:f2supp}c). In this combined network, we show a cobweb plot when we increase $d_1$, and mark the distances $d_i$ as squares on the map (Fig.~\ref{fig:f2supp}d). 
\newpage

\section{Numerically Characterizing Chaos: Lyapunov Exponent}
In the main text, we show a chaotic network module whose map between distances $d_2 = f(d_1)$ is in the form of a general tent map, and which demonstrates chaotic behavior. To quantify this behavior, we use the Lyapunov exponent that measures the sensitivity of the trajectory of an iterated map to minute changes in initial conditions \cite{Strogatz2018Nonlinear}. We iterate the map $n$ times for an initial distance $d_1$ and a small perturbation $d_1' = d_1 + \delta_1$,  such that $d_{n+1} = f^n(d_1)$, and $d_{n+1}' = f^n(d_1')$. Then, we measure how far the trajectories have diverged as $\delta_{n+1} = d_{n+1}' - d_{n+1}$. The Lyapunov exponent captures the exponential rate, $\lambda$, at which this divergence occurs according to $|\delta_{n+1}| = |\delta_1|e^{n\lambda}$. For the trajectory of distances $d_1, d_2,\dotsm$ at the limit of $n \rightarrow \infty$, this exponent is defined as the average log of the slope at each distance
\begin{align*}
\lambda = \lim_{n\rightarrow \infty} \left( \frac{1}{n} \sum_{i=1}^{n} \ln |f'(d_i)| \right).
\end{align*}
Across 1000 evenly spaced $d_1$ across the full map range, we find a tight distribution of positive exponents averaged around $\lambda \approx 0.312 > 0$, which is a hallmark of chaotic systems.
\begin{figure}[h!]
	\centering
	\includegraphics[width=\columnwidth]{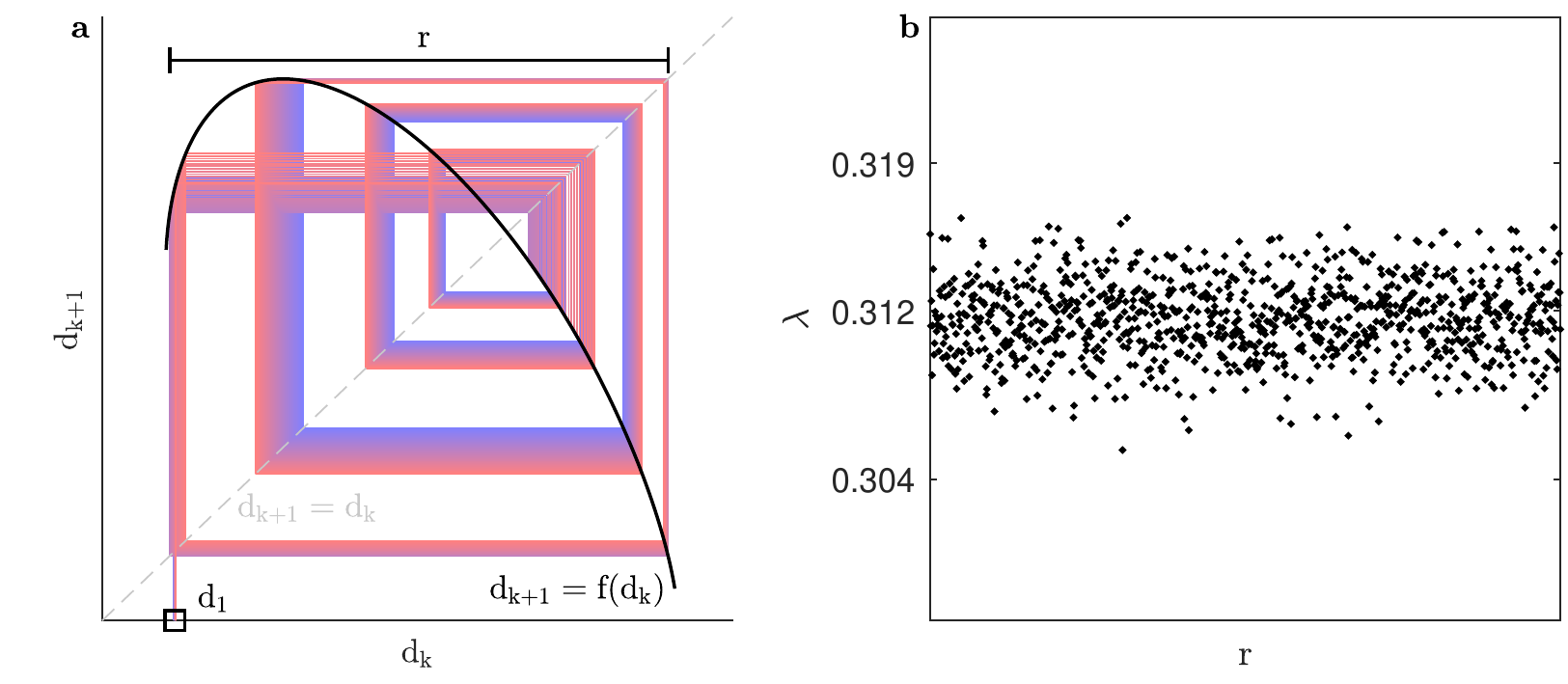}
	\caption{\textbf{Trajectory divergence at chaos.} (\textbf{a}) Map $d_{k+1}=f(d_k)$ of the chaotic module, starting at 50 evenly spaced initial distances from $d_1 = 0.759$ to $d_1 = 0.764$, whose trajectories diverge after 11 steps. The map's domain and range are bounded by a range of distances $r$. (\textbf{b}) We sample 1000 evenly spaced points along $r$ as initial distances, iterate our map 5000 times to settle the distances into the attractor, and compute the Lyapunov exponent for a subsequent 20000 map iterations.}
	\label{fig:f3supp}
\end{figure}
\newpage

\section{Maps of Physical Linkage and 3D-Printed Modules}
\begin{figure}[h!]
	\centering
	\includegraphics[width=\columnwidth]{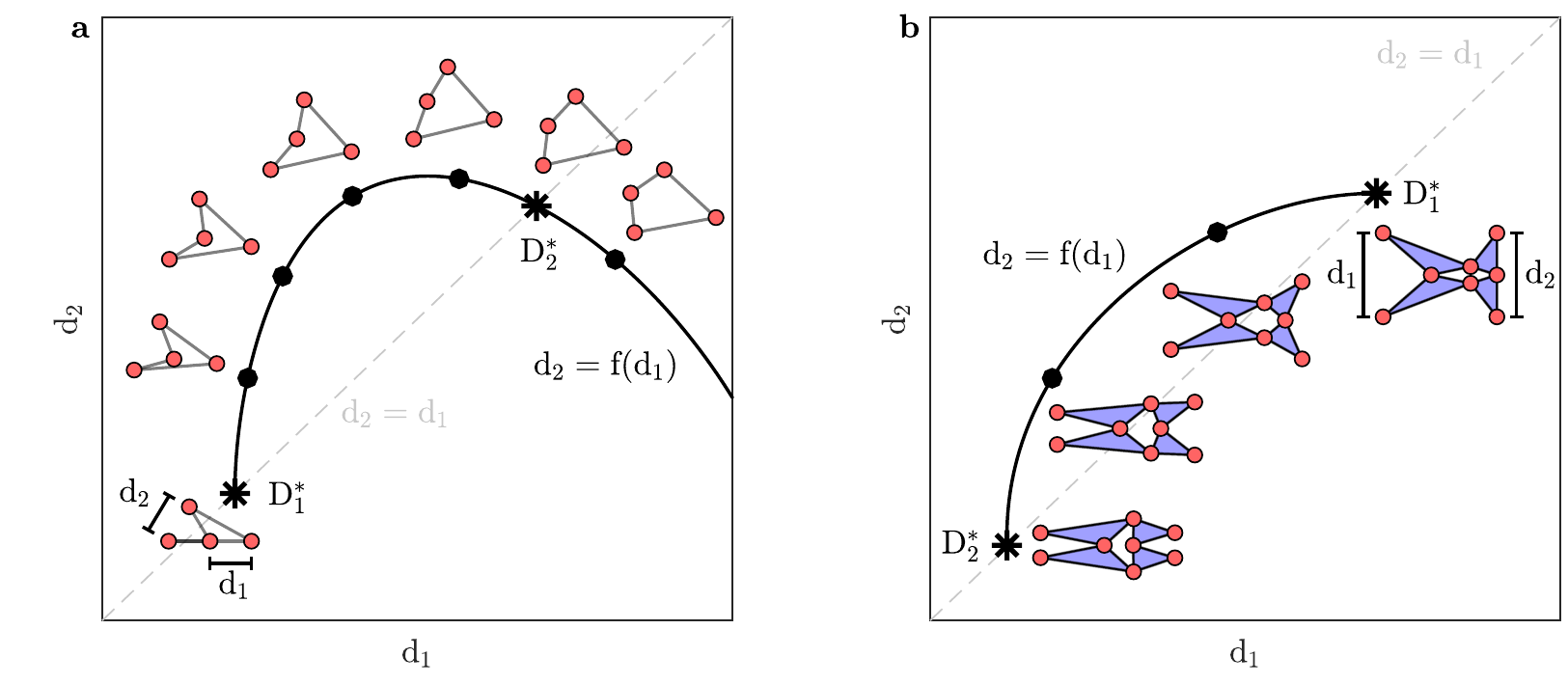}
	\caption{\textbf{Maps of physical networks.} (\textbf{a}) Map $d_2=f(d_1)$ of the 4-bar linkage used as a deployable example, with a super-unstable fixed-point $D_1^*$ and a stable fixed-point $D_2^*$. (\textbf{b}) Map $d_2=f(d_1)$ of the planar network, with a super-stable fixed-point $D_1^*$ and a super-unstable fixed-point $D_2^*$.}
	\label{fig:f4supp}
\end{figure}

In the main text, we construct physical networks using our linkage design framework. In addition, we also show a 4-bar linkage and a 3D printed planar network. Here we show the maps $d_2 = f(d_1)$ for these additional networks. For the 4-bar linkage, we begin in an initial fixed-point collapsed state $D_1^*$ that is super-unstable (slope = $\infty$), where an infinitesimal change in $d_2$ produces no change in $d_1$. Further along the motion, we get another stable fixed point $D_2^*$ (Fig.~\ref{fig:f4supp}a). As for the planar network, we can model each triangular face as a triangular linkage, and plot the map of $d_2 = f(d_1)$ along the conformational motion. This module is super-stable at the initial fixed point $D_1^*$, and is additionally super-unstable at the final fixed point $D_2^*$.
\newpage

\section{Construction and Map of Origami Module}
\begin{figure}[h!]
	\centering
	\includegraphics[width=\columnwidth]{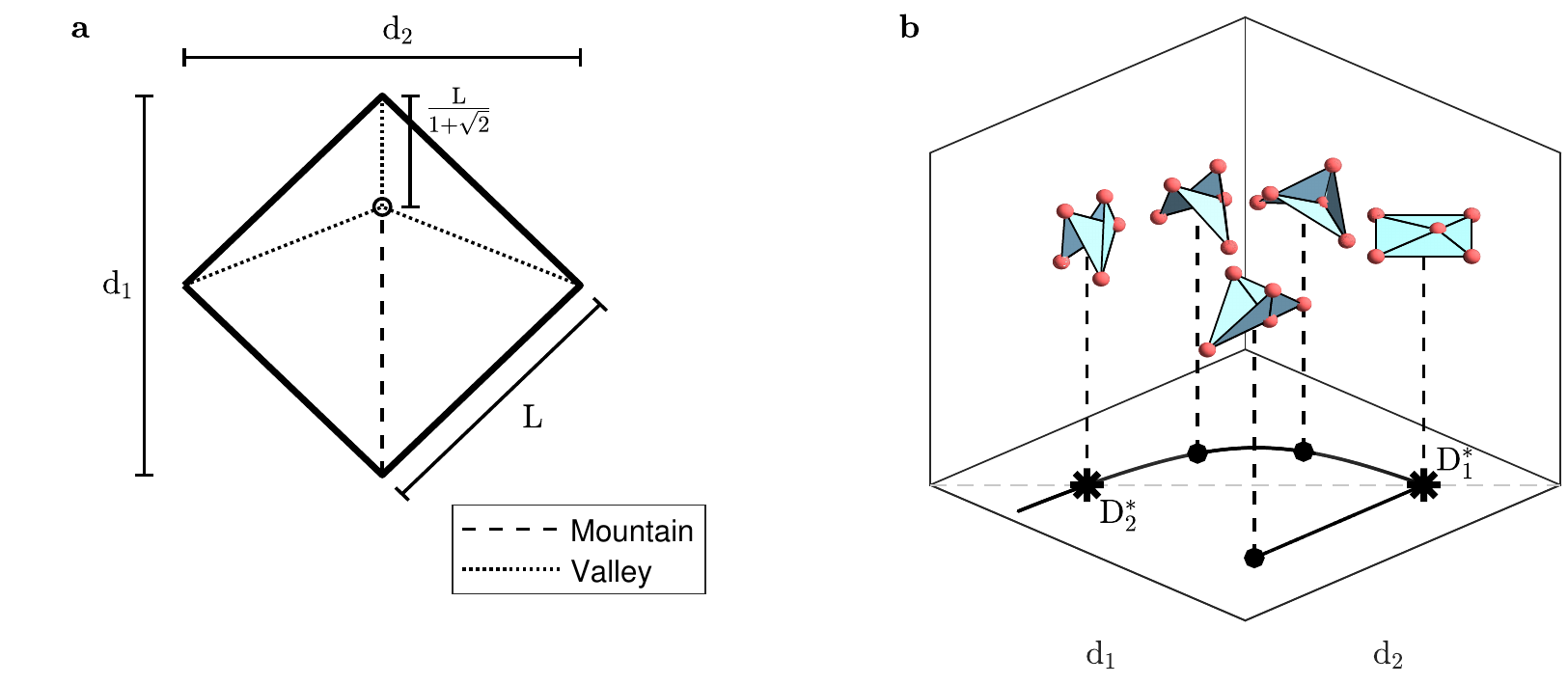}
	\caption{\textbf{Origami sheet construction and map.} (\textbf{a}) Construction of creases in square origami sheet, with labeled dimensions and with mountain and valley folds. We label $d_1$ and $d_2$ as the distance between the corresponding corners. (\textbf{b}) Map $d_2 = f(d_1)$ of this origami sheet that has two fixed points: $D_1^*$ corresponds to the flat sheet configuration, and $D_2^*$ corresponds to a folded configuration. The flat sheet configuration sits at a kinematic bifurcation, where another trajectory exists in which the network simply folds along the main diagonal crease.}
	\label{fig:f5supp}
\end{figure}

Here we detail the construction of the origami module shown in the main text, and numerically plot its map. We begin with a square of paper with side length $L$, and crease it once along the diagonal. We label $d_1$ and $d_2$ as the distances between the opposite marked corners of this square, and also label the mountain and valley creases (Fig.~\ref{fig:f5supp}
a). We then show the map of $d_1$ \emph{versus} $d_2$ along the 1 conformational motion, with the corresponding network geometries shown above sampled points along the map (Fig.~\ref{fig:f5supp}b). At the flat sheet configuration, the network exists at a kinematic bifurcation, such that other possible trajectories exist, such as the network folding along the main diagonal crease with both mountain or valley folds.

\bibliography{references}